\documentclass[12pt,a4paper]{article}
\usepackage{epsfig}
\usepackage{mathrsfs}

\def\hybrid{\topmargin -20pt    \oddsidemargin 0pt
        \headheight 0pt \headsep 0pt
        \textwidth 6.25in       
        \textheight 9.5in       
        \marginparwidth .875in
        \parskip 5pt plus 1pt   \jot = 1.5ex}

\hybrid

\def\baselinestretch{1.2}

\catcode`\@=11

\def\marginnote#1{}
\newcount\hour
\newcount\minute
\newtoks\amorpm
\hour=\time\divide\hour by60
\minute=\time{\multiply\hour by60 \global\advance\minute by-\hour}
\edef\standardtime{{\ifnum\hour<12 \global\amorpm={am}%
        \else\global\amorpm={pm}\advance\hour by-12 \fi
        \ifnum\hour=0 \hour=12 \fi
        \number\hour:\ifnum\minute<10 0\fi\number\minute\the\amorpm}}
\edef\militarytime{\number\hour:\ifnum\minute<10 0\fi\number\minute}

\def\draftlabel#1{{\@bsphack\if@filesw {\let\thepage\relax
   \xdef\@gtempa{\write\@auxout{\string
      \newlabel{#1}{{\@currentlabel}{\thepage}}}}}\@gtempa
   \if@nobreak \ifvmode\nobreak\fi\fi\fi\@esphack}
        \gdef\@eqnlabel{#1}}
\def\@eqnlabel{}
\def\@vacuum{}
\def\draftmarginnote#1{\marginpar{\raggedright\scriptsize\tt#1}}

\def\draft{\oddsidemargin -.5truein
        \def\@oddfoot{\sl preliminary draft \hfil
        \rm\thepage\hfil\sl\today\quad\militarytime}
        \let\@evenfoot\@oddfoot \overfullrule 3pt
        \let\label=\draftlabel
        \let\marginnote=\draftmarginnote
   \def\@eqnnum{(\theequation)\rlap{\kern\marginparsep\tt\@eqnlabel}%
\global\let\@eqnlabel\@vacuum}  }

\def\preprint{\twocolumn\sloppy\flushbottom\parindent 2em
        \leftmargini 2em\leftmarginv .5em\leftmarginvi .5em
        \oddsidemargin -.5in    \evensidemargin -.5in
        \columnsep .4in \footheight 0pt
        \textwidth 10.in        \topmargin  -.4in
        \headheight 12pt \topskip .4in
        \textheight 6.9in \footskip 0pt
        \def\@oddhead{\thepage\hfil\addtocounter{page}{1}\thepage}
        \let\@evenhead\@oddhead \def\@oddfoot{} \def\@evenfoot{} }

\def\numberbysection{\@addtoreset{equation}{section}
        \def\theequation{\thesection.\arabic{equation}}}

\def\underline#1{\relax\ifmmode\@@underline#1\else
        $\@@underline{\hbox{#1}}$\relax\fi}

\def\titlepage{\@restonecolfalse\if@twocolumn\@restonecoltrue\onecolumn
     \else \newpage \fi \thispagestyle{empty}\c@page\z@
        \def\thefootnote{\fnsymbol{footnote}} }

\def\endtitlepage{\if@restonecol\twocolumn \else \newpage \fi
        \def\thefootnote{\arabic{footnote}}
        \setcounter{footnote}{0}}  

\catcode`@=12
\relax

\def\figcap{\section*{Figure Captions\markboth
        {FIGURECAPTIONS}{FIGURECAPTIONS}}\list
        {Figure \arabic{enumi}:\hfill}{\settowidth\labelwidth{Figure
999:}
        \leftmargin\labelwidth
        \advance\leftmargin\labelsep\usecounter{enumi}}}
 \relax
\def\tablecap{\section*{Table Captions\markboth
        {TABLECAPTIONS}{TABLECAPTIONS}}\list
        {Table \arabic{enumi}:\hfill}{\settowidth\labelwidth{Table
999:}
        \leftmargin\labelwidth
        \advance\leftmargin\labelsep\usecounter{enumi}}}
 \relax
\def\reflist{\section*{References\markboth
        {REFLIST}{REFLIST}}\list
        {[\arabic{enumi}]\hfill}{\settowidth\labelwidth{[999]}
        \leftmargin\labelwidth
        \advance\leftmargin\labelsep\usecounter{enumi}}}
 \relax
\makeatletter
\newcounter{pubctr}
\def\publist{\@ifnextchar[{\@publist}{\@@publist}}
\def\@publist[#1]{\list
        {[\arabic{pubctr}]\hfill}{\settowidth\labelwidth{[999]}
        \leftmargin\labelwidth
        \advance\leftmargin\labelsep
        \@nmbrlisttrue\def\@listctr{pubctr}
        \setcounter{pubctr}{#1}\addtocounter{pubctr}{-1}}}
\def\@@publist{\list
        {[\arabic{pubctr}]\hfill}{\settowidth\labelwidth{[999]}
        \leftmargin\labelwidth
        \advance\leftmargin\labelsep
        \@nmbrlisttrue\def\@listctr{pubctr}}}
 \relax
\makeatother
\newskip\humongous \humongous=0pt plus 1000pt minus 1000pt

\newif\ifdtup

\relax

\def\be{\begin{equation}}
\def\ee{\end{equation}}
\def\ba{\begin{eqnarray}}
\def\ea{\end{eqnarray}}

\def\no{\noindent}

\def\IR{\relax{\rm I\kern-.18em R}}

\begin{document}

\renewcommand{\theequation}{\thesection.\arabic{equation}}

\newcommand{\beq}{\begin{equation}}
\newcommand{\eeq}[1]{\label{#1}\end{equation}}
\newcommand{\ber}{\begin{eqnarray}}
\newcommand{\eer}[1]{\label{#1}\end{eqnarray}}
\newcommand{\eqn}[1]{(\ref{#1})}
\begin{titlepage}
\begin{center}

\hfill November 2008\\

\vskip .4in

{\large \bf Duality in linearized gravity and holography}

\vskip 0.6in

{\bf Ioannis Bakas}
\vskip 0.2in
{\em Department of Physics, University of Patras \\
GR-26500 Patras, Greece\\
\footnotesize{\tt bakas@ajax.physics.upatras.gr}}\\

\end{center}

\vskip .8in

\centerline{\bf Abstract}

\no
We consider spherical gravitational perturbations of $AdS_4$ space-time
satisfying general boundary conditions at spatial infinity. Using the
holographic renormalization method, we compute the energy-momentum tensor
and show that it can always be cast in the form of Cotton tensor for a dual
boundary metric. In particular, axial and polar perturbations obeying
the same boundary conditions for the effective Schr\"odinger wave-functions
exhibit an energy-momentum/Cotton tensor duality at conformal infinity.
We demonstrate explicitly that this is holographic manifestation of the
electric/magnetic duality of linearized gravity in the bulk, which simply
exchanges axial with polar perturbations of $AdS_4$ space-time. We note on
the side that this particular realization of gravitational duality is also
valid for perturbations near flat and $dS_4$ space-time, depending on the
value of cosmological constant.

\vfill
\end{titlepage}
\eject

\def\baselinestretch{1.2}
\baselineskip 16 pt
\noindent

\tableofcontents

\section{Introduction}
\setcounter{equation}{0}

AdS/CFT correspondence, \cite{malda}, \cite{igor}, \cite{ed1},
provides a very broad framework to study the relation between
gravitational theories in $d+1$ space-time dimensions and conformal
field theories in $d$ dimensions. Most of the work that has been carried out
so far focused on $d=4$ in connection with supersymmetric Yang-Mills theories
living on the boundary of $AdS_5$ space-time. More recently, there
has been increased interest in aspects of $AdS_4/CFT_3$ correspondence,
whose field theory side is much less understood. The purpose of the
present work is to exploit some special properties of four-dimensional
gravity and obtain their holographic manifestation on the three-dimensional
boundary. Hopefully, this development will help to shed light into the
nature of the boundary field theory and its duality symmetries in
future studies.

We will focus on the simplest case of $AdS_4$ space-time
(without black holes) and derive some properties of the energy-momentum
tensor for general gravitational perturbations satisfying arbitrary
boundary conditions, since they are legitimate to consider by the analysis
of ref. \cite{wald}. It is useful, in this
context, to split the perturbations into two distinct classes and
use the linearized Einstein equations to establish a duality among them.
As will be explained in the sequel, this provides a concrete realization
of the electric/magnetic
duality in linearized gravity, which is rather specific to four space-time
dimensions, and it manifests as energy-momentum/Cotton tensor duality
on the three-dimensional conformal boundary. Setting the boundary free in
$AdS_4/CFT_3$ correspondence, as in ref. \cite{other1}, is the necessary
ingredient in order
to establish these general results. Actually, the existence of a dual
graviton correspondence was anticipated before by some general arguments
in the $AdS_4/CFT_3$ case, \cite{other2}, but a concrete realization
of it was still lacking. Our work fills up this gap and paves the way for
a deeper understanding of the energy-momentum tensor/Cotton tensor duality
for $AdS_4$ black holes that was observed recently, \cite{bakas}.

The realization of the gravitational electric/magnetic duality that will be
obtained along the way has more general value, beyond holography,
since it is totally independent of the size and sign of the cosmological
constant. In fact, it provides for the first time an explicit solution of the
non-local transformation law that connects general metric perturbations
around $(A)dS_4$ and/or Minkowski space-time, depending on the value of
$\Lambda$, which are dual to each other. Our analysis also suggests
possible generalizations of the electric/magnetic duality of linearized
gravity around non-trivial backgrounds, such as the Schwarzschild solution,
which remain to be worked out in detail and applied to the general
theory of gravitational quasi-normal modes with or without cosmological
constant. Thus, the results we present here can be regarded as the simplest
instance of a more general program aiming at the origin of gravitational
duality and its holographic implications, when $\Lambda < 0$.

The material of this paper is organized as follows: In section 2, we
consider the most general spherical perturbations of $AdS_4$ space-time
and derive the equations governing the metric components in the
linear approximation. In section 3, we study the effective Schr\"odinger
equation governing the perturbations and determine the spectrum of
oscillations under general boundary conditions at $r = \infty$.
In section 4, we use the holographic renormalization method to compute
the energy-momentum tensor at the conformal boundary of space-time
for general perturbations satisfying arbitrary boundary conditions.
In section 5, we show that the energy-momentum tensor can be cast in
the form of Cotton tensor for a dual boundary metric, which arises
by exchanging axial with polar perturbations satisfying the same
boundary conditions. In section 6, we show that the holographic
dual graviton correspondence in $AdS_4$ space-time is manifestation of
the electric/magnetic duality of linearized gravity in the bulk that
simply exchanges axial with polar perturbations. Finally, in section
7, we present the conclusions and summarize some open questions and
directions for future work.

\section{Spherical perturbations of $AdS_4$ space-time}
\setcounter{equation}{0}

Einstein equations in four space-time dimensions with
negative cosmological constant,
$R_{\mu \nu} = \Lambda g_{\mu \nu}$, admit $AdS_4$ space-time as solution.
The metric is globally defined using spherical coordinates
\be
ds^2 = -f(r) dt^2 + {dr^2 \over f(r)} + r^2 \left(d\theta^2 +
{\rm sin}^2 \theta d\phi^2 \right)
\ee
with
\be
f(r) = 1 - {\Lambda \over 3} r^2 ~.
\ee

We shall consider perturbations of the metric around the spherically
symmetric static configuration satisfying the linearized Einstein equations
\be
\delta R_{\mu \nu} = \Lambda \delta g_{\mu \nu} ~.
\ee
In four space-time
dimensions, there are two complementary classes of metric perturbations with
opposite parity called axial and polar. In both cases, the equations
reduce to an effective Schr\"odinger problem in the radial direction
and all components of the metric are expressed in terms of the effective
wave-function. The energy levels determine the allowed frequencies of
oscillation around the static solution and their values depend on the
boundary conditions imposed at spatial infinity. The result can be
regarded as specialization of the general theory of quasi-normal modes
of $AdS_4$ black holes in the limit of vanishing mass, $m=0$, so that
global $AdS_4$ space-time is considered instead. Naturally, many
simplifications occur in this case and there are additional features
not to be found when $m \neq 0$. The reader may consult refs. \cite{wheeler},
\cite{zerilli}, \cite{chandra} for the general theory of quasi-normal modes
and refs. \cite{lemos1}, \cite{lemos2}, \cite{moss} for the generalization
to $AdS_4$ black holes.

Let us define for the purposes of the present work the tortoise radial
coordinate $r_{\star}$,
\be
dr_{\star} = {dr \over f(r)} ~,
\ee
so that upon integration we obtain
\be
{\rm tan} \left(\sqrt{-{\Lambda \over 3}} ~ r_{\star} \right) =
\sqrt{-{\Lambda \over 3}} ~ r ~.
\ee
Let us also define for convenience the angular variable
\be
x = \sqrt{-{\Lambda \over 3}} ~ r_{\star} ~,
\ee
which assumes all values from $0$ to $\pi / 2$ as $r$
varies from the origin $r=0$ to spatial infinity $r = \infty$.
Then, the effective Schr\"odinger problem governing the perturbations
of $AdS_4$ space-time assumes the following form,
\be
\left(-{d^2 \over dx^2} + V(x) \right) \Psi (x) = \Omega^2 \Psi (x) ~,
\label{roula}
\ee
for appropriately chosen potential $V(x)$.

Next, we examine separately the two distinct classes of metric perturbations,
$g_{\mu \nu} = g_{\mu \nu}^{(0)} + \delta g_{\mu \nu}$, and tabulate them by
matrices labeled by $(t, r, \theta, \phi)$. Without great loss
of generality, and to simplify the presentation, we only consider axially
symmetric perturbations parametrized by the Legendre polynomials
$P_l({\rm cos} \theta)$; of course, more general perturbations are
expressed in terms of spherical harmonics $Y_l^m (\theta, \phi)$, but
the main results remain essentially the same.

${\bf (i). ~ Axial ~ perturbations:}$
The first class of metric perturbations of $AdS_4$ space-time assume
the following form
\be
\delta g_{\mu \nu} = \left(\begin{array}{cccc}
0 & 0 & 0 & h_0(r) \\
  &   &   &   \\
0 & 0 & 0 & h_1(r) \\
  &   &   &   \\
0 & 0 & 0 & 0 \\
  &   &   &   \\
h_0(r) & h_1(r) & 0 & 0
\end{array} \right)
e^{-i\omega t} {\rm sin} \theta ~
\partial_{\theta} P_l ({\rm cos} \theta) ~,
\ee
The allowed frequencies of oscillation $\omega$ around the equilibrium
configuration will be discussed in the next section.
Axial perturbations correspond to the so called vector sector or shear
channel in the dictionary of AdS/CFT correspondence.

It turns out, without giving all details, that the coefficients of the
metric perturbations are determined in terms of a single function
$\Psi_{\rm RW} (x)$ that satisfies an effective Schr\"odinger equation
\eqn{roula} in the variable $x$ with potential
\be
V_{\rm RW} (x) = {l(l+1) \over {\rm sin}^2 x}
\ee
and
\be
\Omega = \sqrt{-{3 \over \Lambda}} ~ \omega ~.
\label{mouri}
\ee
In particular, Einstein equations amount to the following relations
for the metric functions,
\ba
h_0 (x) & = & {i \over \omega} {d \over dx} \left({\rm tan} x ~
\Psi_{\rm RW} (x) \right) , \\
h_1 (x) & = & \sqrt{-{3 \over \Lambda}} ~ {\rm sin} x ~ {\rm cos} x
~ \Psi_{\rm RW} (x) ~.
\ea

Here, we refer to the quantities of the problem as Regge-Wheeler
potential and wave-functions, using the analogy with the axial
perturbations of four-dimensional black holes, \cite{wheeler}.
Different boundary conditions at spatial infinity, $r = \infty$,
will be encoded into the behavior of the wave-function
$\Psi_{\rm RW} (x)$ at $x= \pi /2$ and specify the solution.

${\bf (ii). ~ Polar ~ perturbations:}$
This is a complementary class of metric perturbations parametrized by
four arbitrary radial functions of the general form
\be
\delta g_{\mu \nu} = \left(\begin{array}{cccc}
f(r)H_0(r) & H_1(r) & 0 & 0 \\
  &   &   &   \\
H_1(r) & H_2(r)/f(r) & 0 & 0 \\
  &   &   &   \\
0 & 0 & r^2K(r) & 0 \\
  &   &   &   \\
0 & 0 & 0 & r^2K(r) {\rm sin}^2 \theta
\end{array} \right)
e^{-i\omega t}
P_l ({\rm cos} \theta)
\ee
They correspond to the so called scalar sector or
sound channel in the dictionary of AdS/CFT correspondence.
The frequencies of oscillation $\omega$ are in general different
for the axial and polar perturbations, depending on the boundary
conditions imposed on each sector.

It turns out, as before, that the coefficients of the
metric perturbations are determined in terms of a single function
$\Psi_{\rm Z} (x)$ that satisfies an effective Schr\"odinger equation
\eqn{roula} in the variable $x$ with potential
\be
V_{\rm Z} (x) = {l(l+1) \over {\rm sin}^2 x}
\ee
that is identical to $V_{\rm RW} (x)$. Also, $\Omega$ is expressed
in terms of the allowed frequencies $\omega$ by equation \eqn{mouri}.
The linearized Einstein equations are satisfied provided that
\be
H_0 (r) = H_2 (r) ~.
\ee
Furthermore, by the same token, the remaining coefficients of the metric
perturbation are expressed in terms $\Psi_{\rm Z} (x)$ as follows,
\be
H_0 (x) = \sqrt{-{\Lambda \over 3}} ~ \left(
{l(l+1) \over 2} {\rm cot} x +
{3 \omega^2 \over \Lambda} {\rm sin} x ~ {\rm cos} x
+ {\rm cos}^2 x {d \over dx} \right) \Psi_{\rm Z} (x) ~,
\ee
\ba
H_1 (x) & = & -i \omega {\rm cos} x ~ {d \over dx} \left({\rm sin} x
~ \Psi_{\rm Z} (x) \right) , \\
K (x) & = & \sqrt{-{\Lambda \over 3}} \left( {l(l+1) \over 2}
{\rm cot} x + {d \over dx} \right) \Psi_{\rm Z} (x) ~.
\ea

Here, we refer to the quantities of the problem as Zerilli
potential and wave-functions, using the analogy with the polar
perturbations of four-dimensional black holes, \cite{zerilli}.
However, unlike the
case of black holes, which exhibit different effective potentials
for the axial and polar perturbations, the perturbations of
$AdS_4$ space-time are governed by the same potential. As before,
different boundary conditions at spatial infinity, $r = \infty$,
will be encoded into the behavior of the wave-function
$\Psi_{\rm Z} (x)$ at $x= \pi /2$.

\section{Boundary conditions and spectrum}
\setcounter{equation}{0}

For both axial and polar perturbations of $AdS_4$ space-time one is led to
consider the effective Schr\"odinger problem
\be
\left(-{d^2 \over dx^2} + {l(l+1) \over {\rm sin}^2 x} \right) \Psi (x)
= \Omega^2 \Psi (x) ~,
\ee
where the frequencies of perturbation are given by
\be
\Omega = \sqrt{-{3 \over \Lambda}} ~ \omega ~.
\ee

This problem can be transformed into a hypergeometric differential equation,
using the change of variables (see, for instance, refs. \cite{wald},
\cite{lemos2})
\be
z= {\rm sin}^2 x ~, ~~~~~ \Psi (x) = {\rm cos} x ~ {\rm sin}^{l+1} x ~
Y(z) ~,
\ee
namely
\be
z(1-z) ~ {d^2 Y \over dz^2} + [c-(a+b+1)z] ~ {d Y \over dz} - ab ~ Y = 0
\ee
with coefficients
\be
a = {1 \over 2} \left(l+2 + \Omega \right) , ~~~~~
b = {1 \over 2} \left(l+2 - \Omega \right)
\ee
and
\be
c = l+ {3 \over 2} ~.
\ee

Then, the solution of the effective Schr\"odinger equation is expressed in
terms of hypergeometric functions (in standard notation) as follows,
\be
\Psi (x) = {\rm cos} x ~ {\rm sin}^{l+1} x ~ F(a, ~ b; ~ c; ~ {\rm sin}^2 x) ~,
\ee
up to an overall numerical factor, so that $\Psi(0) = 0$ at the origin $r=0$.
It provides the normalizable solution of the equation, whereas the other
mathematical solution
\be
\Psi (x) = {{\rm cos} x \over {\rm sin}^{l} x} ~ F(a+1-c, ~ b+1-c; ~ 2- c; ~
{\rm sin}^2 x)
\ee
is not normalizable and blows up at $r=0$. The two solutions have the
familiar $x^{l+1}$ and $x^{-l}$ behavior, respectively, near the origin.
Here, we will only consider the
first one for deriving the energy-momentum tensor of perturbed $AdS_4$
space-time; of course, both solutions should be taken into account to
derive the two-point functions of the energy-momentum tensor, but this
computation is beyond the purpose of the present work.

The behavior of the normalizable solution at spatial infinity follows
by rewriting $\Gamma (a, ~ b; ~ c ; ~ z)$ in terms of the complementary
argument $1-z = {\rm cos}^2 x$. Using standard identities of hypergeometric
functions, it follows that
\ba
\Psi (x) & = & {\Gamma (c) \Gamma (c-a-b)
\over \Gamma (c-a) \Gamma (c-b)} {\rm cos} x ~ {\rm sin}^{l+1} x ~
F(a, ~b; ~ a+b+1-c; ~ {\rm cos}^2 x) + \nonumber \\
& & {\Gamma (c) \Gamma (a+b-c)
\over \Gamma (a) \Gamma (b)} {\rm sin}^{l+1} x ~
F(c-a, ~c-b; ~ 1+c-a-b; ~ {\rm cos}^2 x) ~.
\ea
The wave-function can be expanded in powers of $1/r$, as
\be
\Psi (r) = I_0 + {I_1 \over r} + {I_2 \over r^2} + {I_3 \over r^3}
+ {I_4 \over r^4} + \cdots ~,
\ee
and the first two coefficient turn out to be
\ba
I_0 & = & \Gamma^{-1} \left({1 \over 2} (l+2 + \Omega)\right)
\Gamma^{-1} \left({1 \over 2} (l+2 - \Omega)\right)  \\
I_1 & = & -2 \sqrt{-{3 \over \Lambda}} ~
\Gamma^{-1} \left({1 \over 2} (l+1 + \Omega)\right)
\Gamma^{-1} \left({1 \over 2} (l+1 - \Omega)\right) ,
\ea
up to an overall (irrelevant) numerical factor.

The remaining coefficients of the asymptotic expansion of $\Psi$ in powers
of $1/r$ can be easily determined as
\ba
I_2 & = & -{3I_0 \over 2 \Lambda} \left(l(l+1) + {3 \omega^2 \over \Lambda}
\right) , \\
I_3 & = & -{I_1 \over 2 \Lambda} \left((l-1)(l+2) + {3 \omega^2 \over \Lambda}
\right) , \\
I_4 & = & {3 I_0 \over 8 \Lambda^2} \Big[ \left(l(l+1) +
{3 \omega^2 \over \Lambda}
\right)^2 - 6 \left(l(l+1) + {4 \omega^2 \over \Lambda} \right) \Big]
\ea
and so on. They will be need later (up to the order shown above) for the
computation of the energy-momentum tensor.

The boundary conditions at spatial infinity $r = \infty$ are solely expressed
in terms of $I_0$ and $I_1$. Since
\be
I_0 = \Psi (r= \infty) ~, ~~~~~
{\Lambda \over 3} I_1 = {d \Psi \over d r_{\star}} (r = \infty) ~,
\ee
it follows that general boundary conditions (also called mixed or Robin)
can be expressed in terms of the ratio
\be
{I_0 \over I_1} = \gamma
\ee
for fixed constant $\gamma$ that can assume all values, including zero and
infinity. Thus, the allowed spectrum of frequencies $\omega$ obeys a
transcendental relation given by ratios of gamma functions and can only
be solved numerical for general values of $\gamma$. In all cases, however,
the frequencies come in pairs $(\omega, - \omega)$, as can be readily seen
from the particular expressions of $I_0$ and $I_1$ in terms of products of
gamma functions. Consequently, by appropriate superposition of them, the
metric perturbations can always be taken real.

There are two special boundary conditions that yield simple spectrum of
perturbations. First, we consider Dirichlet boundary conditions, which are
favored in most applications in AdS/CFT correspondence, by imposing
$I_0 = 0$. Since the gamma function blows up only when its argument
assumes the values $0, -1, -2, \cdots $, it follows that the quantization
condition for the frequencies is
\be
\sqrt{-{3 \over \Lambda}} ~ \omega_{\rm D} = \pm (2n + l + 2)
\ee
with $n= 0, 1, 2, \cdots $.
Second, we may also consider Neumann boundary conditions by imposing
$I_1 = 0$. In this case, the corresponding quantization condition for the
frequencies takes the form
\be
\sqrt{-{3 \over \Lambda}} ~ \omega_{\rm N} = \pm (2n + l + 1)
\ee
with $n= 0, 1, 2, \cdots $.
Note that for given $l$, the frequencies that yield $I_0 = 0$
have $I_1 \neq 0$ and those that yield $I_1 = 0$ have $I_0 \neq 0$.
We also note that $\omega_{\rm N}$ follows from $\omega_{\rm D}$ by
letting $l \rightarrow l-1$.

The effective Schr\"odinger potential derives from a superpotential
$W(x)$,
\be
W(x) = l ~ {\rm cot} x ~,
\ee
since
\be
V_{\mp}(x) = W^2(x) \mp {d \over dx} W(x) = {l(l \pm 1) \over
{\rm sin}^2 x} - l^2 ~.
\ee
These are supersymmetric partner potentials which happen to follow from
each other by the simple substitution $l \rightarrow l-1$.
The energy levels $E$ of $V_{\mp}(x)$ provide the allowed
frequencies of perturbation,
\be
E= -{3 \over \Lambda} ~ \omega^2 - l^2 ~,
\ee
and the corresponding wave-functions $\Psi_{\mp} (x)$ with the same $E$
(and hence the same $\omega$) are interrelated as
\be
\left(\mp {d \over dx} + W(x) \right) \Psi_{\pm} (x) =
\sqrt{E} ~  \Psi_{\mp} (x)
\ee
by supersymmetric quantum mechanics, \cite{susy}. In turn, this implies,
setting $x=\pi /2$, that
\be
\mp {d \over dx} \Psi_{\pm} (\pi /2) = \sqrt{E} ~ \Psi_{\mp} (\pi /2) ~,
\ee
exchanging Dirichlet and Neumann boundary conditions under
$l \rightarrow l-1$ for the particular Schr\"odinger problem. This
explains the relation among $\omega_{\rm D}$ and $\omega_{\rm N}$
noted earlier.

More generally, mixed boundary conditions with parameter $\gamma$,
as defined above, are mapped to mixed boundary conditions with
parameter
\be
\gamma \rightarrow {\Lambda \over 3E} \cdot {1 \over \gamma}
\ee
under $l \rightarrow l-1$, so that a given frequency $\omega$
appears in the spectra of both $V_{\mp}(x)$.
This relation is also apparent from the particular
form of the coefficients $I_0$ and $I_1$ above, using the identity
$\Gamma (z+1) = z \Gamma (z)$, but has no special meaning in the
holographic description of the perturbed space-time, as far as
we can tell now.

For fixed $l$, the gravitational perturbations of $AdS_4$ space-time
can satisfy arbitrary boundary conditions at spatial infinity. This
possibility was investigated in the literature before, following the
thorough analysis of ref. \cite{wald}, and it was found that all
boundary conditions are allowed as they are in one-to-one correspondence
with the self-adjoint extensions of the positive definite operator
\be
L = -{d^2 \over dx^2} + {l(l+1) \over {\rm sin}^2 x} ~.
\ee

Summarizing, the boundary conditions for axial and polar perturbations
of $AdS_4$
space-time are independent from each other, and so is the spectrum of
allowed frequencies, whereas the effective Schr\"odinger equations
are the same. Thus, for fixed $l$, the axial and polar perturbations
are isospectral provided that they both satisfy the same boundary
conditions. It should be contrasted to the situation of $AdS_4$
black holes whose axial and polar perturbations obey supersymmetric
partner potential Schr\"odinger equations for any given $l$,
\cite{lemos1}. In the latter case, axial and polar perturbations will be
isospectral if and only if the corresponding boundary conditions
are supersymmetric partners.

\section{Energy-momentum tensor}
\setcounter{equation}{0}

The energy-momentum tensor on asymptotically AdS spaces $M$ is defined
by varying the gravitational action $S_{\rm gr}$ with respect to the
boundary metric $\gamma$ on $\partial M$, as
\be
T^{ab} = {2 \over \sqrt{-{\rm det} \gamma}} {\delta S_{\rm gr} \over
\delta \gamma_{ab}} ~.
\ee
The resulting
expression typically diverge, but it is always possible by holographic
renormalization to obtain
finite results adding an appropriately chosen boundary counter-term
whose form depends on the dimensionality of space-time, \cite{skenderis1},
\cite{skenderis2}, \cite{kraus}.

In $AdS_4$ spaces, in particular, the gravitational action
consists of bulk and boundary terms chosen as follows,
\ba
S_{\rm gr} & = & -{1 \over 2 \kappa^2} \int_{M} d^4 x \sqrt{-{\rm det} g}
\left(R[g] + 2 \Lambda \right) - {1 \over \kappa^2}
\int_{\partial M} d^3 x \sqrt{-{\rm det} \gamma} ~ K \nonumber\\
& & - {2 \over \kappa^2} \sqrt{-{\Lambda \over 3}} \int_{\partial M} d^3 x
\sqrt{-{\rm det} \gamma} \left(1 + {3 \over 4 \Lambda} R[\gamma]
\right) .
\ea
The first boundary contribution is the usual Gibbons-Hawking term written
in terms of the trace of the second fundamental form, i.e., the
extrinsic mean curvature, $K = \gamma^{ab} K_{ab}$,
associated to the embedding of $\partial M$ in $M$. The second boundary
contribution is the contact term needed to remove all divergencies in
the present case.
Then, according to definition, the energy-momentum tensor
is expressed in terms of the intrinsic and extrinsic
geometry of the AdS boundary at infinity, prior to rescaling, as
\be
\kappa^2 T_{ab} = K_{ab} - K \gamma_{ab} -2 \sqrt{-{\Lambda \over 3}}
\gamma_{ab} + \sqrt{-{3 \over \Lambda}} \left(
R_{ab}[\gamma] - {1 \over 2} R[\gamma] \gamma_{ab} \right) ~.
\ee

The computation is performed by first writing the metric
$g$ on $M$ in the form
\be
ds^2 = N^2 dr^2 + \gamma_{ab} \left(dx^a + N^a dr \right)
\left(dx^b + N^b dr \right)
\ee
using appropriately chosen $(N, N^a)$ functions.
The three-dimensional surface arising at fixed distance
$r$ serves as boundary $\partial M_r$ to the interior
four-dimensional region $M_r$. The induced metric on $\partial M_r$
is $\gamma_{ab}$ evaluated at the boundary value of $r$, which is
held finite at this point.
The second fundamental form $K_{ab}$ on $\partial M_r$ is defined
using the outward pointing normal vector $\eta_{\mu}$ to the
boundary $\partial M_r$ with components $\eta_{\mu} = N ~ \delta_{\mu}^r$.
In particular, one has
\be
K_{ab} = - \nabla_{(a} \eta_{b)} = N ~ \Gamma_{ab}^r [g] ~.
\ee
At the end of the computation, $T^{ab}$ on the AdS boundary $\partial M$
is obtained by letting $r \rightarrow \infty$.

The boundary metric acquires an infinite Weyl factor as $r$ is taken to infinity,
and, therefore, it is more appropriate to think of the AdS boundary
as a conformal class of boundaries and define $\mathscr{I}$ as the boundary
space-time with metric
\be
ds_{\mathscr{I}}^2 = \lim_{r \rightarrow \infty} \left(-{3 \over \Lambda r^2}
\gamma_{ab} dx^a dx^b \right) .
\ee
Then, the renormalized energy-momentum tensor on $\mathscr{I}$ is defined
accordingly by
\be
T_{ab}^{\rm renorm} = \lim_{r \rightarrow \infty} \left(\sqrt{-{\Lambda \over 3}}
~ r ~ T_{ab} \right)
\ee
and it is finite, traceless and conserved.
This is the quantity that will be computed for all different type
of gravitational perturbations of $AdS_4$ space-time. The decorations will be
dropped in the sequel to simplify the notation.

${\bf (i). ~ Axial ~ perturbations:}$ Applying the holographic renormalization
method to axial perturbations of $AdS_4$ space-time is straightforward. The
boundary data depends on the coefficients $I_0$ and $I_1$ in the asymptotic
expansion of the effective wave-function in powers of $1/r$,
\be
\Psi_{\rm RW} (r) = I_0 + {I_1 \over r} + {I_2 \over r^2} + {I_3 \over r^3}
+ \cdots ~,
\ee
since all subleading terms are fixed uniquely by the choice of boundary
conditions. Since the calculation is quite involved, we only present the final
result for the energy-momentum tensor and metric at the conformal boundary
when arbitrary boundary conditions are imposed at spatial infinity.

The three-dimensional metric on $\mathscr{I}$
takes the following form, after conformal rescaling,
\be
ds_{\mathscr{I}}^2 = -dt^2 -{3 \over \Lambda} \left(d \theta^2 +
{\rm sin}^2 \theta d \phi^2 \right) + 2 {iI_0 \over \omega}
e^{-i \omega t} {\rm sin} \theta ~ \partial_{\theta} P_l ({\rm cos} \theta)
~ dt d \phi
\ee
and the non-vanishing components of the renormalized energy-momentum tensor
for axial perturbations of $AdS_4$ space-time turn out to be
\ba
\kappa^2 T_{t \phi} & = & { i \Lambda \over 6 \omega}
(l-1)(l+2) I_1 e^{-i \omega t} {\rm sin} \theta ~
\partial_{\theta} P_l({\rm cos} \theta) ~, \\
\kappa^2 T_{\theta \phi} & = &
- {1 \over 2} I_1 e^{-i \omega t} {\rm sin} \theta [l(l+1)
~ P_l ({\rm cos} \theta) + 2 {\rm cot} \theta ~
\partial_{\theta} P_l ({\rm cos} \theta)] ~.
\ea

It can be easily verified that this
energy-momentum tensor is traceless and conserved on $\mathscr{I}$.
It vanishes only when $I_1 = 0$, i.e., when $\Psi_{\rm RW}$
satisfies Neumann boundary conditions at spatial infinity. At the same
time, the boundary metric depends on time and it becomes static only
when $I_0 = 0$, i.e., when $\Psi_{\rm RW}$
satisfies Dirichlet boundary conditions at spatial infinity.

${\bf (ii). ~ Polar ~ perturbations:}$ The computation of the energy-momentum
tensor for polar perturbations of $AdS_4$ space-time is much more involved.
It requires one more term in the asymptotic expansion of $\Psi_{\rm Z}$,
\be
\Psi_{\rm Z} (r) = J_0 + {J_1 \over r} + {J_2 \over r^2} + {J_3 \over r^3}
+ {J_4 \over r^4} + \cdots ~,
\ee
but all boundary data is fully determined by $J_0$ and $J_1$ as before.
Here, we use the symbols $J_i$ to distinguish from the symbols $I_i$
appearing as coefficients in the asymptotic expansion of the axial
wave-functions. We
also skip the intermediate details and present the final result for the
energy-momentum tensor and metric at the conformal boundary
when arbitrary boundary conditions are imposed at spatial infinity.

In this case, the
three-dimensional metric on the boundary takes the following form, after
conformal rescaling,
\be
ds_{\mathscr{I}}^2 = -dt^2 -{3 \over \Lambda} \Big[1+ {\Lambda \over 3} J_1
e^{-i \omega t}
P_l ({\rm cos} \theta) \Big] (d\theta^2 + {\rm sin}^2 \theta d \phi^2 ) ~.
\ee

Explicit calculation shows that the non-vanishing components of the
renormalized energy-momentum tensor for polar perturbations of $AdS_4$
space-time are:
\ba
\kappa^2 T_{tt} & = & - {\Lambda \over 12} (l-1)l (l+1)(l+2) J_0
e^{-i \omega t} P_l ({\rm cos} \theta) ~, \\
\kappa^2 T_{\theta \theta} & = & - {1 \over 4} l(l+1) \left(1 +
{3 \omega^2 \over \Lambda} \right) J_0 e^{-i \omega t}
P_l({\rm cos} \theta) \nonumber\\
& & - {1 \over 4} \left(l(l+1)
+ {6 \omega^2 \over \Lambda} \right) J_0 e^{-i \omega t}
{\rm cot} \theta ~ \partial_{\theta} P_l({\rm cos} \theta) , \\
\kappa^2 T_{\phi \phi} & = & {1 \over 4} l(l+1) \left(l(l+1) - 1 +
{3 \omega^2 \over \Lambda} \right) J_0 e^{-i \omega t}
{\rm sin}^2 \theta P_l({\rm cos} \theta) + \nonumber\\
& & {1 \over 4} \left(l(l+1)
+ {6 \omega^2 \over \Lambda} \right) J_0
e^{-i \omega t} {\rm sin} \theta {\rm cos} \theta ~
\partial_{\theta} P_l({\rm cos} \theta) ~, \\
\kappa^2 T_{t \theta} & = & {1 \over 4} i \omega
(l-1)(l+2) J_0 e^{-i \omega t} \partial_{\theta} P_l({\rm cos} \theta) ~.
\ea

It can be verified, as consistency check, that the energy-momentum
tensor is traceless and conserved on $\mathscr{I}$.
It vanishes only when $J_0 = 0$, i.e., when $\Psi_{\rm Z}$
satisfies Dirichlet boundary conditions at spatial infinity. At the same
time, the boundary metric depends on time and it becomes static only
when $J_1 = 0$, i.e., when $\Psi_{\rm Z}$
satisfies Neumann boundary conditions at spatial infinity. This behavior is
complementary to the axial perturbations studied above and calls for an
explanation.

\section{Cotton tensor duality}
\setcounter{equation}{0}

When the boundary metric is static it is also conformally flat, since it
corresponds to $2+1$ dimensional Einstein universe
\be
ds_{\mathscr{I}}^2 = -dt^2 -{3 \over \Lambda} \left(d \theta^2 +
{\rm sin}^2 \theta d \phi^2 \right) ~.
\ee
Otherwise, for general boundary conditions, the boundary metric varies with time.
A way to measure the deviation from the conformally flat form is provided by the
Weyl tensor of the metric, but in three space-time dimensions all components
vanish identically. In this case, however, there is an alternative way provided
by the so called Cotton tensor of the metric.

The Cotton tensor of a three-dimensional metric $\gamma_{ab}$ is defined as follows,
\ba
C^{ab} & = & {1 \over 2 \sqrt{- {\rm det} \gamma}} \left( \epsilon^{acd}
\nabla_c {R^b}_d + \epsilon^{bcd} \nabla_c {R^a}_d \right) \nonumber\\
& = & {\epsilon^{acd} \over \sqrt{- {\rm det} \gamma}} \nabla_c
\left({R^b}_d - {1 \over 4} {\delta^b}_d R \right)
\ea
and has odd parity. The density $\sqrt{{\rm det} \gamma } ~ {C^a}_b$
remains invariant under local conformal changes and vanishes if and only if the
metric is conformally flat. The Cotton tensor is symmetric, traceless and
identically covariantly conserved. It provides the energy-momentum tensor of
the three-dimensional gravitational Chern-Simons theory,
\cite{cs},
\be
C_{ab} = {1 \over \sqrt{- {\rm det} \gamma}}
{\delta S_{\rm CS} \over \delta \gamma^{ab}}
\ee
with action
\be
S_{\rm CS} = {1 \over 2} \int d^3 x \sqrt{- {\rm det} \gamma} ~
\epsilon^{abc} \Gamma_{ae}^d \left(
\partial_b \Gamma_{cd}^e + {2 \over 3} \Gamma_{bf}^e \Gamma_{cd}^f \right) .
\ee

Let us now compute the Cotton tensor for the boundary metric of perturbed
$AdS_4$ space-time, setting $\epsilon^{t \theta \phi} = 1$. We study separately
the axial and polar perturbations satisfying arbitrary boundary conditions.

${\bf (i). ~ Axial ~ perturbations:}$ In this case, the boundary metric takes
the form
\be
ds_{\mathscr{I}}^2 ({\rm axial}) = -dt^2 -{3 \over \Lambda} \left(d \theta^2 +
{\rm sin}^2 \theta d \phi^2 \right) + 2 {iI_0 \over \omega}
e^{-i \omega t} {\rm sin} \theta ~ \partial_{\theta} P_l ({\rm cos} \theta)
~ dt d \phi
\ee
when general boundary conditions with coefficients $I_0$ and $I_1$ are imposed
at spatial infinity on the axial wave-function.

Then, explicit computation shows that the non-vanishing components of the
corresponding Cotton tensor are
\ba
C_{tt} & = & {i \Lambda^2 \over 18 \omega} (l-1)l (l+1)(l+2) I_0
e^{-i \omega t} P_l ({\rm cos} \theta) ~, \\
C_{\theta \theta} & = & {i \Lambda \over 6 \omega} l(l+1) \left(1 +
{3 \omega^2 \over \Lambda} \right) I_0 e^{-i \omega t}
P_l({\rm cos} \theta) + \nonumber\\
& & {i \Lambda \over 6 \omega} \left(l(l+1)
+ {6 \omega^2 \over \Lambda} \right) I_0 e^{-i \omega t}
{\rm cot} \theta ~ \partial_{\theta} P_l({\rm cos} \theta) , \\
C_{\phi \phi} & = & -{i \Lambda \over 6 \omega} l(l+1) \left(l(l+1) - 1 +
{3 \omega^2 \over \Lambda} \right) I_0 e^{-i \omega t}
{\rm sin}^2 \theta P_l({\rm cos} \theta)  \nonumber\\
& & - {i \Lambda \over 6 \omega} \left(l(l+1)
+ {6 \omega^2 \over \Lambda} \right) I_0
e^{-i \omega t} {\rm sin} \theta {\rm cos} \theta ~
\partial_{\theta} P_l({\rm cos} \theta) ~, \\
C_{t \theta} & = & {\Lambda \over 6}
(l-1)(l+2) I_0 e^{-i \omega t} \partial_{\theta} P_l({\rm cos} \theta) ~.
\ea
and coincide with the components of the energy-momentum tensor for polar
perturbations of $AdS_4$ space-time satisfying general boundary conditions
with respective coefficients $J_0$ and $J_1$, provided that
\be
J_0 = -i {2 \Lambda \over 3 \omega} I_0 ~.
\ee

The match is exact provided that the same frequencies $\omega$ arise
for the axial and polar perturbations, i.e., when the same boundary
conditions are imposed at spatial infinity,
\be
{I_0 \over I_1} = {J_0 \over J_1} ~.
\label{meri}
\ee
Then, we have the following relation among the two distinct type of
perturbations satisfying the same general boundary conditions,
\be
C_{ab} ({\rm axial}) = \kappa^2 T_{ab} ({\rm polar}) ~.
\label{roi1}
\ee

${\bf (ii). ~ Polar ~ perturbations:}$ In this case, the boundary metric takes
the form
\be
ds_{\mathscr{I}}^2 ({\rm polar})
= -dt^2 -{3 \over \Lambda} \Big[1+ {\Lambda \over 3} J_1 e^{-i \omega t}
P_l ({\rm cos} \theta) \Big] (d\theta^2 + {\rm sin}^2 \theta d \phi^2 ) ~.
\ee
when general boundary conditions with coefficients $J_0$ and $J_1$ are imposed
at spatial infinity on the polar wave-function.

Again, explicit computation shows that the non-vanishing components of the
corresponding Cotton tensor are
\ba
C_{t \phi} & = & {\Lambda^2 \over 36}
(l-1)(l+2) J_1 e^{-i \omega t} {\rm sin} \theta ~
\partial_{\theta} P_l({\rm cos} \theta) ~, \\
C_{\theta \phi} & = &
{\Lambda \over 12} i \omega J_1 e^{-i \omega t} {\rm sin} \theta [l(l+1)
~ P_l ({\rm cos} \theta) + 2 {\rm cot} \theta ~
\partial_{\theta} P_l ({\rm cos} \theta)] ~.
\ea
and coincide with the components of the energy-momentum tensor for axial
perturbations of $AdS_4$ space-time satisfying general boundary conditions
with respective coefficients $I_0$ and $I_1$, provided that
\be
I_1 = - {\Lambda \over 6} i \omega J_1 ~.
\ee

As before, the match is exact provided that the same frequencies
$\omega$ arise for the axial and polar perturbations, i.e., when the same
boundary conditions \eqn{meri} are imposed on both sectors at spatial
infinity. Then, under this condition, we obtain the following general
relation connecting the two perturbations,
\be
C_{ab} ({\rm polar}) = \kappa^2 T_{ab} ({\rm axial}) ~.
\label{roi2}
\ee

Note at this end that the situation changes drastically in the presence
of black holes in $AdS_4$ space-time, \cite{bakas}.
First of all, the holographic
energy-momentum tensor of the $AdS_4$ Schwarzschild solution does not
vanish, as its components depend on the mass parameter. Also,
according to the general theory of
quasi-normal modes in four space-time dimensions, axial and polar
perturbations of the Schwarzschild solution satisfy different Schr\"odinger
equations, which are usually referred as Regge-Wheeler and Zerilli
equations, respectively. However, the two effective potentials are
partners as in supersymmetric quantum mechanics, \cite{susy}, and, as
a result, there is still a relation between the energy-momentum tensor
of perturbed black holes and the Cotton tensor of a boundary dual metric
as given by
\be
C_{ab} ({\rm axial}) = \kappa^2 \delta T_{ab} ({\rm polar}) ~, ~~~~~
C_{ab} ({\rm polar}) = \kappa^2 \delta T_{ab} ({\rm axial}) ~.
\ee
It was noted before, \cite{bakas}, that this relation is only valid when
the axial and polar perturbations satisfy specific supersymmetric partner
boundary conditions that encompass the hydrodynamic modes of black holes.
The result should be contrasted with the energy-momentum/Cotton tensor
duality for perturbed $AdS_4$ space-time, which, as described above,
is valid for all boundary conditions provided that they are the same in
both sectors.

\section{Electric/magnetic duality in linearized gravity}
\setcounter{equation}{0}

Free gauge fields in four space-time dimensions exhibit two physical degrees
of freedom that can be rotated into one another by a canonical transformation
mixing the two pairs of unconstrained dynamical variables, while
keeping the Hamiltonian form-invariant. It is a general result that extends the
electric/magnetic duality of electromagnetism to other physical fields
including Einstein gravity. For gravity, the duality is defined at the linear
level by considering small perturbations around a reference metric,
$g_{\mu \nu} = g_{\mu \nu}^{(0)} + \delta g_{\mu \nu}$.
There is a non-local transformation to perturbations around the same
reference metric $\tilde{g}_{\mu \nu} =
g_{\mu \nu}^{(0)} + \delta \tilde{g}_{\mu \nu}$,  acting on the space of
solutions of vacuum Einstein equation and which can also be
realized as symmetry of the gravitational action. It generalizes the
Ehlers transformation\footnote{The space of vacuum solutions of Einstein
field equations with a Killing isometry in four space-time dimensions admit
the action of an $SL(2, R)$ group known as Ehlers symmetry, \cite{ehlers}.
This group acts as solution generating symmetry and it is present at the
full non-linear level only when $\Lambda = 0$. It arises by rewriting the
four-dimensional Einstein-Hilbert action as three-dimensional gravity
coupled to an $SL(2, R)/U(1)$ non-linear sigma model. An $SO(2)$
subgroup acts continuously as $S$-duality and interchanges
``bolts" and ``nuts" in space-time as in electric/magnetic duality,
\cite{gary} (but see also ref. \cite{tst}).
The Ehlers symmetry breaks down in the presence of cosmological
constant, unlike the electric/magnetic duality of linearized gravity that
holds for all values of $\Lambda$.}
to metrics that do not necessarily admit Killing isometries, but it breaks
down at first self-interacting cubic approximation to general relativity,
\cite{stan2}, which, however, is not relevant to the present work.
The reference metric $g_{\mu \nu}^{(0)}$ is not arbitrary in this study;
it is provided by the metric of flat Minkowski space-time
when the cosmological constant $\Lambda$ vanishes and by the metric of
$(A)dS_4$ space-time when $\Lambda \neq 0$.

The electric/magnetic duality of linearized gravity was initially formulated
for $\Lambda = 0$, \cite{claudio}, but it extends rather easily to vacuum
Einstein equations with cosmological constant, \cite{stan1}, \cite{julia},
\cite{tassos}; see also ref. \cite{nieto} for earlier important work on the same
subject.  Here, we choose to work with $\Lambda < 0$, although
the description below is valid for all values of the cosmological constant.
The key quantity is provided by
\be
Z_{\mu \nu \rho \sigma} = R_{\mu \nu \rho \sigma} - {\Lambda \over 3}
\left(g_{\mu \rho} g_{\nu \sigma} - g_{\mu \sigma} g_{\nu \rho} \right)
\ee
that arises by restricting the Weyl curvature tensor in four space-time
dimensions to on-shell metrics, \cite{julia}. Clearly, it fulfills the identities
\be
Z_{\mu [\nu \rho \sigma ]} = 0 ~, ~~~~~
\nabla_{[ \lambda} Z_{\mu \nu ] \rho \sigma} = 0 ~,
\ee
and the on-shell metrics satisfy the equation
\be
{Z^{\rho}}_{\mu \rho \nu} \equiv Z_{\mu \nu} = 0 ~,
\ee
which is equivalent to $R_{\mu \nu} = \Lambda g_{\mu \nu}$.
One also defines the dual curvature tensor
\be
\tilde{Z}_{\mu \nu \rho \sigma} = {1 \over 2}
{\epsilon_{\mu \nu}}^{\kappa \lambda} Z_{\kappa \lambda \rho \sigma} ~,
\ee
which fulfills similar identities, but with reverse meaning,
\be
\tilde{Z}_{\mu [\nu \rho \sigma ]} = 0 ~, ~~~~~
\nabla_{[ \lambda} \tilde{Z}_{\mu \nu ] \rho \sigma} = 0
\ee
and it satisfies the classical equation of motion
\be
{\tilde{Z}^{\rho}}_{~ \mu \rho \nu} \equiv \tilde{Z}_{\mu \nu} = 0 ~.
\ee
Here, ${\epsilon_{\mu \nu}}^{\kappa \lambda}$ is the covariant
fully antisymmetric symbol in four-dimensional space-time with
$\epsilon_{r t \theta \phi} = \sqrt{-{\rm det} g}$.

Linearized gravity around $AdS_4$ space-time exhibits a duality that
exchanges Bianchi identities with the classical equations of motion,
as in electromagnetism, by
\be
Z_{\mu \nu \rho \sigma}^{\prime} = \tilde{Z}_{\mu \nu \rho \sigma} ~,
~~~~~
\tilde{Z}_{\mu \nu \rho \sigma}^{\prime} = - Z_{\mu \nu \rho \sigma} ~.
\ee
Actually, at the linear level, it is appropriate to replace the covariant
derivatives $\nabla$ by ordinary derivatives $\partial$.
Also, it is useful to introduce the electric and magnetic components
of the Weyl tensor as
\be
{\cal E}_{ab} = Z_{a r b r} ~, ~~~~~ {\cal B}_{ab} =
\tilde{Z}_{a r b r} ~,
\ee
using the radial coordinate $r$ of $AdS_4$ space-time.
Then, the gravitational duality transformation is realized as
\be
{\cal E}_{ab}^{\prime} = {\cal B}_{ab} ~, ~~~~~
{\cal B}_{ab}^{\prime} = - {\cal E}_{ab} ~,
\label{dual}
\ee
and, more generally, one may consider an $SO(2)$ rotation of these
components parametrized by an arbitrary angle $\delta$,
\be
\left(\begin{array}{c}
{\cal E}_{ab}^{\prime} \\
                       \\
{\cal B}_{ab}^{\prime}
\end{array} \right) =
\left(\begin{array}{ccc}
{\rm cos} \delta &  & {\rm sin} \delta \\
                 &  &                \\
- {\rm sin} \delta  &  & {\rm cos} \delta
\end{array} \right)
\left(\begin{array}{c}
{\cal E}_{ab} \\
                       \\
{\cal B}_{ab}
\end{array} \right) .
\ee
Details of the proof can be found in the original works showing that
the duality is also a symmetry of the linearized action (see, in
particular, ref. \cite{claudio} and \cite{julia}).

The electric and magnetic tensors are represented by $3 \times 3$
symmetric traceless matrices on-shell, and, as such, they have five
independent components each. Let us compute them for the axial and
polar perturbations of $AdS_4$ space-time and show that their
interchange accounts for the duality \eqn{dual}.

${\bf (i). ~ Axial ~ perturbations:}$ Taking into account the general
form of the axial perturbations around $AdS_4$ space-time and their
explicit dependence on the frequencies $\omega$ and the wave-functions
$\Psi_{\rm RW}$, we find the following results on-shell for the electric
components
\ba
{\cal E}_{\theta \phi}^{\rm axial} & = & -{{\rm cos}^3 x \over 2}
{d \over dx} \left({\rm sin} x ~ \Psi_{\rm RW} (x) \right) e^{-i \omega t}
{\rm sin} \theta ~ [(l(l+1) + 2 {\rm cot} \theta
~ \partial_{\theta}] P_{l} ({\rm cos} \theta) ~, \\
{\cal E}_{t \phi}^{\rm axial} & = & {i \Lambda \over 6 \omega}
(l-1)(l+2) ~ {{\rm cos}^3 x \over {\rm sin} x} \left({d \over dx}
\Psi_{\rm RW} (x) \right) e^{-i \omega t} {\rm sin} \theta ~
\partial_{\theta} P_{l} ({\rm cos} \theta) ~,
\ea
and similarly for the magnetic components
\ba
{\cal B}_{tt}^{\rm axial} & = & {i \over 2 \omega} \left(\sqrt{-{\Lambda
\over 3}} \right)^3 (l-1) l (l+1) (l+2) {{\rm cos}^3 x \over {\rm sin}^3 x}
\Psi_{\rm RW} (x) e^{-i \omega t} P_{l} ({\rm cos} \theta) ~, \\
{\cal B}_{\theta \theta}^{\rm axial} & = & -{i \over 2 \omega} \sqrt{-
{\Lambda \over 3}} ~ l(l+1) {{\rm cos}^3 x \over {\rm sin} x} \Big[ \left(
1 + {3 \omega^2 \over \Lambda} {\rm sin}^2 x \right) \Psi_{\rm RW} (x)
+ \nonumber\\
& & ~~~ {\rm sin} x ~ {\rm cos} x {d \over dx} \Psi_{\rm RW} (x) \Big]
e^{-i \omega t} P_{l} ({\rm cos} \theta) \nonumber\\
& & - {i \over \omega} \sqrt{-
{\Lambda \over 3}} ~ {{\rm cos}^3 x \over {\rm sin} x} \Big[ \left(
{l(l+1) \over 2} + {3 \omega^2 \over \Lambda} {\rm sin}^2 x \right)
\Psi_{\rm RW} (x) + \nonumber\\
& & ~~~ {\rm sin} x ~ {\rm cos} x {d \over dx} \Psi_{\rm RW} (x) \Big]
e^{-i \omega t} {\rm cot} \theta ~ \partial_{\theta}
P_{l} ({\rm cos} \theta) ~, \\
{\cal B}_{\phi \phi}^{\rm axial} & = &  {i \over 2 \omega} \sqrt{-
{\Lambda \over 3}} ~ l(l+1) {{\rm cos}^3 x \over {\rm sin} x}
\Big[ \left(l^2 + l - 1 + {3 \omega^2 \over \Lambda} {\rm sin}^2 x \right)
\Psi_{\rm RW} (x) + \nonumber\\
& & ~~~ {\rm sin} x ~ {\rm cos} x {d \over dx} \Psi_{\rm RW} (x) \Big]
e^{-i \omega t} {\rm sin}^2 \theta ~ P_{l} ({\rm cos} \theta) \nonumber\\
& & + {i \over \omega} \sqrt{-
{\Lambda \over 3}} ~ {{\rm cos}^3 x \over {\rm sin} x} \Big[ \left(
{l(l+1) \over 2} + {3 \omega^2 \over \Lambda} {\rm sin}^2 x \right)
\Psi_{\rm RW} (x) + \nonumber\\
& & ~~~ {\rm sin} x ~ {\rm cos} x {d \over dx} \Psi_{\rm RW} (x) \Big]
e^{-i \omega t} {\rm sin} \theta {\rm cos} \theta ~ \partial_{\theta}
P_{l} ({\rm cos} \theta) ~, \\
{\cal B}_{t \theta}^{\rm axial} & = & -{1 \over 2} \sqrt{-
{\Lambda \over 3}} ~ (l-1) (l+2) {{\rm cos}^3 x \over {\rm sin} x}
\Psi_{\rm RW} (x) e^{-i \omega t}
\partial_{\theta} P_{l} ({\rm cos} \theta) ~.
\ea

Here, we have also used that the metric of axially perturbed space-time has
determinant
\be
\sqrt{-{\rm det} g} = r^2 {\rm sin} \theta = -{3 \over \Lambda} {\rm tan}^2 x
~ {\rm sin} \theta ~.
\ee

${\bf (ii). ~ Polar ~ perturbations:}$ Performing the same calculation
for the polar perturbations of $AdS_4$ space-time, using their
explicit dependence on the corresponding frequencies $\omega$ and the
wave-functions $\Psi_{\rm Z}$, we find the following results on-shell
for the electric components
\ba
{\cal E}_{tt}^{\rm polar} & = & - {1 \over 4} \left(\sqrt{-{\Lambda
\over 3}} \right)^3 (l-1) l (l+1) (l+2) {{\rm cos}^3 x \over {\rm sin}^3 x}
\Psi_{\rm Z} (x) e^{-i \omega t} P_{l} ({\rm cos} \theta) ~, \\
{\cal E}_{\theta \theta}^{\rm polar} & = & {1 \over 4} \sqrt{-
{\Lambda \over 3}} ~ l(l+1) {{\rm cos}^3 x \over {\rm sin} x} \Big[ \left(
1 + {3 \omega^2 \over \Lambda} {\rm sin}^2 x \right) \Psi_{\rm Z} (x)
+ \nonumber\\
& & ~~~ {\rm sin} x ~ {\rm cos} x {d \over dx} \Psi_{\rm Z} (x) \Big]
e^{-i \omega t} P_{l} ({\rm cos} \theta) \nonumber\\
& & + {1 \over 2} \sqrt{-
{\Lambda \over 3}} ~ {{\rm cos}^3 x \over {\rm sin} x} \Big[ \left(
{l(l+1) \over 2} + {3 \omega^2 \over \Lambda} {\rm sin}^2 x \right)
\Psi_{\rm Z} (x) + \nonumber\\
& & ~~~ {\rm sin} x ~ {\rm cos} x {d \over dx} \Psi_{\rm Z} (x) \Big]
e^{-i \omega t} {\rm cot} \theta ~ \partial_{\theta}
P_{l} ({\rm cos} \theta) ~, \\
{\cal E}_{\phi \phi}^{\rm polar} & = &  - {1 \over 4} \sqrt{-
{\Lambda \over 3}} ~ l(l+1) {{\rm cos}^3 x \over {\rm sin} x}
\Big[ \left(l^2 + l - 1 + {3 \omega^2 \over \Lambda} {\rm sin}^2 x \right)
\Psi_{\rm Z} (x) + \nonumber\\
& & ~~~ {\rm sin} x ~ {\rm cos} x {d \over dx} \Psi_{\rm Z} (x) \Big]
e^{-i \omega t} {\rm sin}^2 \theta ~ P_{l} ({\rm cos} \theta) \nonumber\\
& & - {1 \over 2} \sqrt{-
{\Lambda \over 3}} ~ {{\rm cos}^3 x \over {\rm sin} x} \Big[ \left(
{l(l+1) \over 2} + {3 \omega^2 \over \Lambda} {\rm sin}^2 x \right)
\Psi_{\rm Z} (x) + \nonumber\\
& & ~~~ {\rm sin} x ~ {\rm cos} x {d \over dx} \Psi_{\rm Z} (x) \Big]
e^{-i \omega t} {\rm sin} \theta {\rm cos} \theta ~ \partial_{\theta}
P_{l} ({\rm cos} \theta) ~, \\
{\cal E}_{t \theta}^{\rm polar} & = & -{i \omega \over 4} \sqrt{-
{\Lambda \over 3}} ~ (l-1) (l+2) {{\rm cos}^3 x \over {\rm sin} x}
\Psi_{\rm Z} (x) e^{-i \omega t}
\partial_{\theta} P_{l} ({\rm cos} \theta) ~,
\ea
and similarly for the magnetic components
\ba
{\cal B}_{\theta \phi}^{\rm polar} & = & i \omega {{\rm cos}^3 x \over 4}
{d \over dx} \left({\rm sin} x ~ \Psi_{\rm Z} (x) \right) e^{-i \omega t}
{\rm sin} \theta ~ [(l(l+1) + 2 {\rm cot} \theta
~ \partial_{\theta}] P_{l} ({\rm cos} \theta) ~, \\
{\cal B}_{t \phi}^{\rm polar} & = & {\Lambda \over 12}
(l-1)(l+2) ~ {{\rm cos}^3 x \over {\rm sin} x} \left({d \over dx}
\Psi_{\rm Z} (x) \right) e^{-i \omega t} {\rm sin} \theta ~
\partial_{\theta} P_{l} ({\rm cos} \theta) ~.
\ea

Here, the determinant of the metric of perturbed space-time is given by
\be
\sqrt{-{\rm det} g} = r^2 {\rm sin} \theta \left(1 + K(r) e^{-i \omega t}
P_l ({\rm cos} \theta) \right) ,
\ee
which, in turn, can be written in terms of the angular variable $x$,
instead of $r$, used in the computations.

Since both perturbations satisfy the same Schr\"odinger equation, we
have $\Psi_{\rm RW} = \Psi_{\rm Z}$ (up to an arbitrary multiplicative
constant) when the same boundary conditions
are imposed at infinity; they also ensure that the frequencies $\omega$
are the same in both cases. In fact, choosing the multiplicative
constant as
\be
\Psi_{\rm RW} (x) = {i \omega \over 2} ~ \Psi_{\rm Z} (x)
\ee
the result of the calculation is neatly summarized as
\be
{\cal E}_{ab}^{\rm polar} = {\cal B}_{ab}^{\rm axial} , ~~~~~
{\cal B}_{ab}^{\rm polar} = - {\cal E}_{ab}^{\rm axial}
\ee
showing that the gravitational duality \eqn{dual} is realized by
exchanging axial with polar perturbations, as advertised above.

This simple fact has not been spelled in the literature so far,
to the best of
our knowledge, since gravitational duality was only considered as
abstract transformation acting non-locally on the perturbations
$\delta g_{\mu \nu}$. It is also obviously valid for all values of
the cosmological constant. Recall, for this purpose, that
$x = r_{\star} \sqrt{- \Lambda / 3}$ and, therefore, the effective
Schr\"odinger problem takes the following form in the limit $\Lambda = 0$,
\be
\left(-{d^2 \over dr^2} + {l(l+1) \over r^2} \right) \Psi (r) =
\omega^2 \Psi (r)
\ee
in terms of the radial variable $r$ ($=r_{\star}$), instead of $x$.
Then, ${\cal E}_{ab}$ and ${\cal B}_{ab}$
assume their respective values obtained from the $\Lambda$-dependent
expressions above, which are well defined and involve $\Psi (r)$ and
its derivatives with respect to $r$. Thus, for $\Lambda = 0$, the
duality of linearized gravity is also realized by interchanging axial and
polar perturbations of flat space-time, without ever using the form of
the wave-functions $\Psi (r)$. Likewise, for perturbations
of $dS_4$ space-time, one simply has to consider the analytic
continuation of the trigonometric functions of $x$ into their
hyperbolic counterparts and reach the same conclusion. The exchange of
axial and polar perturbations resembles the exchange of ``nuts" and
``bolts" under the action of Ehlers symmetry, where it is appropriate
to use.

The holographic manifestation in $AdS_4$ space-time is
the energy-momentum/Cotton tensor duality at the conformal boundary
that gives rise to the dual graviton correspondence. This is made
more precise by further noticing
\be
\lim_{r \rightarrow \infty} \left({\Lambda \over 3} r^3 {\cal E}_{ab}
\right) = \kappa^2 T_{ab} ~, ~~~~
\lim_{r \rightarrow \infty} \left({\Lambda^2 \over 9} r^3 {\cal B}_{ab}
\right) = C_{ab}
\ee
for either type of perturbation, using the asymptotic expansion of the
effective Schr\"odinger wave-functions, and completes the proof of the
main assertion for perturbations of $AdS_4$.

\section{Conclusions}
\setcounter{equation}{0}

We have obtained a precise realization of electric/magnetic duality in
linearized gravity in terms of the interchange of axial and polar
perturbations around $AdS_4$ space-time. Since a generic perturbation
can be decomposed into a sum of axial and polar parts, our results
``trivialize" in a certain sense the action of dualities.
We have also shown that the holographic manifestation of this duality is
the energy-momentum/Cotton tensor
duality that realizes the dual graviton correspondence in $AdS_4/CFT_3$
under general boundary conditions.
These results should be taken into account in future work to investigate the
structure of the three-dimensional field theory at the conformal boundary.

When a black hole is added in space-time the situation changes. It is not
known whether the duality of linearized gravity persists for perturbations
of the Schwarzschild solution. Still one has two distinct type of
perturbations, axial and polar, as in $AdS_4$ space-time, but they do not
satisfy the same effective Schr\"odinger equation. They rather satisfy
supersymmetric partner Schr\"odinger equations, irrespective of $\Lambda$,
as noted before.
It is natural to expect that this partnership among axial and polar
perturbations can also be understood in terms of gravitational duality
when appropriately formulated on black hole backgrounds. This
possibility is currently under investigation and hopefully will explain
why supersymmetric quantum mechanics is at work in the four-dimensional
theory of quasi-normal modes of black holes for all values of $\Lambda$.
Perhaps that is the best one can do for perturbations of non-trivial
gravitational backgrounds. In turn, it
may also provide a deeper understanding of the energy-momentum/Cotton
tensor duality for perturbations of $AdS_4$ black holes with appropriate
boundary conditions on the axial and polar sectors of the correspondence,
\cite{bakas}.

We also note that there are special configurations (other than
the cases we are considering here) which satisfy the self-duality relations
\be
\tilde{Z}_{\mu \nu \rho \sigma} (g) = \pm Z_{\mu \nu \rho \sigma}
(\tilde{g})
\ee
with $g_{\mu \nu} = \tilde{g}_{\mu \nu}$. They define the so called
$\Lambda$-instantons, in the nomenclature of ref. \cite{julia}, whose
energy-momentum tensor equals the Cotton tensor of their boundary metric
$\gamma_{ab}$, when $\Lambda < 0$, \cite{feffe} (but see also ref.
\cite{petkou}),
\be
\kappa^2  T_{ab} ({\rm instantons}) = C_{ab} (\gamma) ~.
\ee
It will be interesting to examine the fate of gravitational duality
rotations for perturbations around the metric of $\Lambda$-instantons - not
just around global $AdS_4$ space-time - and find their boundary
manifestation by holographic techniques. We hope to return to this problem
elsewhere.

Finally, it will be very interesting to explore the role of gravitational
duality
in the two-point functions of the energy-momentum tensor for spherical
perturbations of $AdS_4$ space-time, and more generally of $AdS_4$
Schwarzschild solution, using the results presented here as well as those
reported in ref. \cite{bakas}. Work in this direction is in progress,
\cite{baske}.

\newpage

\centerline{\bf Acknowledgements}

This work was supported in part by the European Research and Training Network
``Constituents, Fundamental Forces and Symmetries of the Universe"
(MRTN-CT-2004-005104) and by the bilateral research grant
``Gravity, Gauge Theories and String Theory"
(06FR-020) for Greek-French scientific cooperation.
I also thank the theory groups at CERN, NORDITA and the Arnold Sommerfeld
Center for hospitality and financial
support during the course of the present work.


\newpage

\begin{thebibliography}{3}

\bibitem{malda}
J. Maldacena, ``The large $N$ limit of superconformal
field theories and supergravity", Adv. Theor. Math. Phys.
\underline{2} (1998) 231 [hep-th/9711200].
\bibitem{igor}
S.S. Gubser, I.R. Klebanov and A.M. Polyakov, ``Gauge theory correlators
from noncritical string theory", Phys. Lett. \underline{B428} (1998) 105
[hep-th/9802109].
\bibitem{ed1}
E. Witten, ``Anti-de Sitter space and holography", Adv. Theor. Math. Phys.
\underline{2} (1998) 253 [hep-th/9802150].
\bibitem{wald}
A. Ishibashi and R.M. Wald, ``Dynamics on non-globally-hyperbolic static
spacetimes: III. Anti-de Sitter spacetime", Class. Quant. Grav.
\underline{21} (2004) 2981 [hep-th/0402184].
\bibitem{other1}
G. Compere and D. Marolf, ``Setting the boundary free in AdS/CFT", Class. Quant.
Grav. \underline{25} (2008) 195014 [arXiv:0805.1902].
\bibitem{other2}
S. de Haro, ``Dual gravitons in $AdS_4/CFT_3$ and the holographic Cotton tensor"
[arXiv:0808.2054].
\bibitem{bakas}
I. Bakas, ``Energy-momentum/Cotton tensor duality for $AdS_4$ black holes"
[arXiv:0809.4852].
\bibitem{wheeler}
T. Regge and J.A. Wheeler, ``Stability of a Schwarzschild singularity",
Phys. Rev. \underline{108} (1957) 1063.
\bibitem{zerilli}
F.J. Zerilli, ``Effective potential for even-parity Regge-Wheeler gravitational
perturbation equations", Phys. Rev. Lett. \underline{24} (1970) 737.
\bibitem{chandra}
S. Chandrasekhar, {\em The Mathematical Theory of Black Holes},
Oxford University Press, Oxford, 1983.
\bibitem{lemos1}
V. Cardoso and J.P.S. Lemos, ``Quasinormal modes of Schwarzschild anti-de Sitter
black holes: Electromagnetic and gravitational perturbations", Phys. Rev.
\underline{D64} (2001) 084017 [gr-qc/0105103].
\bibitem{lemos2}
V. Cardoso, R. Konoplya and
J.P.S. Lemos, ``Quasinormal frequencies of Schwarzschild black holes in
anti-de Sitter space-times: A complete study on the overtone asymptotic
behavior", Phys. Rev. \underline{D68} (2003) 044024 [gr-qc/0305037].
\bibitem{moss}
I.G. Moss and J.P. Norman, ``Gravitational quasinormal modes for anti-de
Sitter black holes", Class. Quant. Grav. \underline{19} (2002) 2323
[gr-qc/0201016].
\bibitem{susy}
F. Cooper, A. Khare and U. Sukhatme, ``Supersymmetry and quantum mechanics",
Phys. Rept. \underline{251} (1995) 267 [hep-th/9405029].
\bibitem{skenderis1}
S. de Haro, S.N. Solodukhin and K. Skenderis, ``Holographic reconstruction
of space-time and renormalization in AdS/CFT correspondence", Commun. Math.
Phys. \underline{217} (2001) 595
[hep-th/0002230].
\bibitem{skenderis2}
K. Skenderis, ``Asymptotically anti-de Sitter space-times and their stress
energy tensor", Int. J. Mod. Phys. \underline{A16} (2001) 740 [hep-th/0010138];
``Lecture notes on holographic renormalization", Class. Quant. Grav.
\underline{19} (2002) 5849
[hep-th/0209067].
\bibitem{kraus}
V. Balasubramanian and P. Kraus, ``A stress tensor for anti-de Sitter gravity",
Commun. Math. Phys. \underline{208} (1999) 413 [hep-th/9902121].
\bibitem{cs}
S. Deser, R. Jackiw and S. Templeton, ``Topologically massive gauge theories",
Ann. Phys. \underline{140} (1982) 372; Erratum-ibid. \underline{185}
(1988) 406; ``Three-dimensional massive gauge theories", Phys. Rev. Lett.
\underline{48} (1982) 975.
\bibitem{ehlers}
J. Ehlers, ``Transformations of static exterior solutions of Einstein's
gravitational field equations into different solutions by means of conformal
mapping" in {\em Les Th\'eories Relativistes de la Gravitation},
Colloq. Int. CNRS \underline{91} (1961) 275.
\bibitem{gary}
G.W. Gibbons and S.W. Hawking, ``Classification of gravitational instanton
symmetries", Commun. Math. Phys. \underline{66} (1979) 291.
\bibitem{tst}
I. Bakas, ``Space-time interpretation of $S$-duality and supersymmetry
violations of $T$-duality", Phys. Lett. \underline{B343} (1995) 103
[hep-th/9410104].
\bibitem{stan2}
S. Deser and D. Seminara,  ``Free spin 2 duality invariance cannot be extended
to GR", Phys. Rev. \underline{D71} (2005) 081502 [hep-th/0503030].
\bibitem{claudio}
M. Henneaux and C. Teitelboim, ``Duality in linearized gravity", Phys. Rev.
\underline{D71} (2005) 024018 [gr-qc/0408101].
\bibitem{stan1}
S. Deser and D. Seminara, ``Duality invariance of all free bosonic and
fermionic gauge fields", Phys. Lett. \underline{B607} (2005) 317
[hep-th/0411169].
\bibitem{julia}
B. Julia, J. Levie and S. Ray, ``Gravitational duality near de Sitter space",
JHEP \underline{0511} (2005) 025 [hep-th/0507262]; B. Julia,
``Electric-magnetic duality beyond four dimensions and in general relativity",
talk at the {\em 23rd International Conference on Differential Geometric
Methods in Theoretical Physics}, Tianjin, China [hep-th/0512320].
\bibitem{tassos}
R.G. Leigh and A.C. Petkou, ``Gravitational duality transformations on
$(A)dS_4$", JHEP \underline{0711} (2007) 079 [arXiv:0704.0531].
\bibitem{nieto}
J.A. Nieto, ``S-duality for linearized gravity", Phys. Lett. \underline{A262}
(1999) 274 [hep-th/9910049].
\bibitem{feffe}
C. Fefferman and C.R. Graham, ``The ambient metric" [arXiv:0710.0919].
\bibitem{petkou}
D.S. Mansi, A.C. Petkou and G. Tagliabue, ``Gravity in the $3+1$-split formalism
II: Self-duality and the emergence of the gravitational Chern-Simons in the
boundary" [arXiv:0808.1213].
\bibitem{baske}
I. Bakas and K. Skenderis, work in progress.

\end{thebibliography}
\end{document}